\documentclass[a4paper,preprintnumbers,floatfix,superscriptaddress,pra,twocolumn,showpacs,notitlepage,longbibliography]{revtex4-1}
\usepackage[utf8]{inputenc}
\usepackage[T1]{fontenc}
\usepackage[sc,osf]{mathpazo}
\usepackage{amsmath, amsthm, amssymb,amsfonts,mathbbol,amstext}
\usepackage{graphicx}
\usepackage{dcolumn}
\usepackage{bm}
\usepackage{bbm}
\usepackage{hyperref}
\usepackage{mathtools}
\usepackage{comment}
\usepackage{color}
\usepackage{dsfont}
\usepackage[normalem]{ulem}
\usepackage{multirow}
\usepackage{diagbox}
\usepackage{amsmath}
\usepackage{amsthm}
\usepackage{bbm}
\usepackage{comment} 
\usepackage{marvosym}
\usepackage{graphicx}
\usepackage{xcolor}
\usepackage{appendix}
\usepackage{hyperref}

\newtheorem{theorem}{Theorem}

\newtheorem{result}[theorem]{Result}

\definecolor{applegreen}{rgb}{0.55, 0.71, 0.0}
\definecolor{darkelectricblue}{rgb}{0.03, 0.51, 0.57}
\definecolor{atomictangerine}{rgb}{1.0, 0.6, 0.4}

\newcommand{\ket}[1]{| #1 \rangle}
\newcommand{\bra}[1]{\langle #1 |}

\newcommand{\stkout}[1]{\ifmmode\text{\sout{\ensuremath{#1}}}\else\sout{#1}\fi}

\definecolor{forestgreen}{rgb}{0.13, 0.55, 0.13}

\definecolor{sam}{rgb}{0.5, 0., 0.5}

\DeclareMathOperator*{\maximize}{Maximize} 

\def\Id{\mathbbm{1}}

 \DeclareMathOperator{\tr}{tr}
\DeclareMathOperator{\rank}{rank}

\begin{document}
\title{Semi-device-independent certification of entanglement in superdense coding}
\author{George Moreno}
\affiliation{International Institute of Physics, Federal University of Rio Grande do Norte, 59070-405 Natal, Brazil}
\author{Ranieri Nery}
\affiliation{International Institute of Physics, Federal University of Rio Grande do Norte, 59070-405 Natal, Brazil}
\author{Carlos de Gois}
\affiliation{Instituto  de  Física ``Gleb  Wataghin'', Universidade  Estadual  de  Campinas, CEP 13083-859, Campinas, Brazil}
\author{Rafael Rabelo}
\affiliation{Instituto  de  Física ``Gleb  Wataghin'', Universidade  Estadual  de  Campinas, CEP 13083-859, Campinas, Brazil}
\author{Rafael Chaves}
\affiliation{International Institute of Physics, Federal University of Rio Grande do Norte, 59070-405 Natal, Brazil}
\affiliation{School of Science and Technology, Federal University of Rio Grande do Norte, 59078-970 Natal, Brazil}

\date{\today}
\begin{abstract}

Superdense coding is a paradigmatic protocol in quantum information science, employing a quantum communication channel to send classical information more efficiently. As we show here, it can be understood as a particular case of a prepare and measure experiment, a scenario that has attracted growing attention for its fundamental and practical applications. Formulating superdense coding as a prepare and measure scenario allows us to provide a semi-device-independent witness of entanglement that significantly improves over previous tests. Furthermore, we also show how to adapt our results into self-testing of maximally entangled states and also provide a semidefinite program formulation allowing one to efficiently optimize, for any shared quantum state, the probability of success in the superdense coding protocol.
\end{abstract}

\maketitle
\section{Introduction}
Quantum communication \cite{Gisin2007} is arguably among the first offsprings of quantum technologies to break out of the laboratory. Recent milestones, such as quantum teleportation using metropolitan networks \cite{Valivarthi2016} and satellites sharing entanglement across continental and intercontinental distances \cite{Yin2017,Liao2018}, are paving the way for the realistic implementation of many of the quantum communication protocols discovered over the last years. Of particular relevance is the possibility of large scale quantum networks, the so-called quantum internet \cite{wehner2018quantum,brito2020statistical}, not only allowing for more efficient communication \cite{Bennett1992,Bennett93} but also for fundamental information security \cite{Gisin2002}.

In such applications it is of utmost importance to be able to certify the nonclassicality of the quantum resources, typically the presence of quantum entanglement \cite{Horodecki2009} between the communicating parties. For instance, entangled states allow for better teleported states \cite{Cavalcanti2017}, improved communication efficiency in the superdense coding protocol \cite{Bennett1992} and quantum cryptography \cite{Bennett92QKD}. However, in order to detect any quantum enhancement in these examples, one needs to have full control over the preparation as well as of the measurement apparatuses. In practice, noise is unavoidable, potentially leading to erroneous conclusions \cite{rosset2012} and opening the way to hacker attacks \cite{lydersen2010hacking}. To cope with that, the device-independent (DI) framework has been established \cite{brunner2014bell}. Based on mild general assumptions, it allows one to certify quantumness simply from the observational data, not requiring any detailed knowledge of the underlying physical mechanisms at play.

The DI framework emerged in the context of Bell's theorem \cite{Bell1964}, finding use in practical applications ranging from quantum key distribution \cite{Ekert1991,acin2007device,vazirani2019fully} to communication complexity \cite{buhrman2010nonlocality} and self-testing \cite{mayers2003self,vsupic2020self}. In spite of its clear importance, however, the Bell scenario turns out to be rather restrictive in the context of quantum communication, since only pre-established correlations but no communication are allowed. More recently, device-independent scenarios allowing for communication have started to attract growing attention. Of particular relevance is the so-called prepare and measure (PAM) scenario, a fairly general structure that, apart from its foundational relevance \cite{pawlowski2009information,chaves2015information,chaves2018causal}, has found applications in quantum networks \cite{bowles2015testing,wang2019characterising}, self-testing \cite{tavakoli2018self,miklin2020universal}, quantum key distribution \cite{pawlowski-qkd-2011}, randomness certification \cite{passaro-randomness-2015}, random access codes \cite{li2012semi} and as nonclassicality witnesses \cite{gallego-pam-2010,chaves2015device,van2017semi,tavakoli2020characterising,tavakoli2020informationally,poderini2020criteria}. Apart from exploratory attempts in \cite{pawlowski2010entanglement}, in all these works the communicating parties share classical correlations; the nonclassicality can only arise due to the communicating (nonentangled) quantum states. As a consequence, the PAM scenario and the kind of device-independence it entails have not yet found any use in the most relevant entanglement-enhanced quantum communication protocols. That is precisely the problem we solve here.

As we show, the paradigmatic superdense coding \cite{Bennett1992} can be cast as a particular instance of the prepare and measure scenario. As a consequence, a dimension witness quantifying the probability of success of the superdense coding \cite{Brunner2013} can also be used to certify, in a semi-DI manner, the nonclassicality of the shared correlations between the communicating parties. As opposed to the typical Bell scenario that is fully DI, quantum communication scenarios have to impose a limit on the amount of communication exchanged, otherwise the communication problem becomes trivial. In line with the superdense coding protocol, we achieve that by imposing a limit on the Hilbert-space dimension of the quantum system being communicated. Strikingly, no other information about the preparation and measurement devices is required. Nicely, any pure bipartite entangled state as well as a large family of entangled mixed states violate our witness. Our results largely improve over other semi-DI witnesses of entanglement: not only do they reduce the experimental requirements and increase the tolerance to noise, but they also do not require partial state tomography to work, such as in quantum steering \cite{Wiseman2007}. We also provide a semidefinite program (SDP) formulation allowing one to obtain lower bounds for the optimal probability of superdense coding success for arbitrary shared states. Following that we show how the nonclassicality in the superdense coding naturally leads to a self-testing protocol, also discussing its limitations in cryptographic scenarios. Finally, we also go beyond the superdense coding, analyzing a more general prepare and measure scenario allowing for quantum correlations and a measurement device with several inputs.

\section{Superdense coding as a prepare and measure scenario}
\begin{figure}
    \centering
    \includegraphics[scale = 0.25]{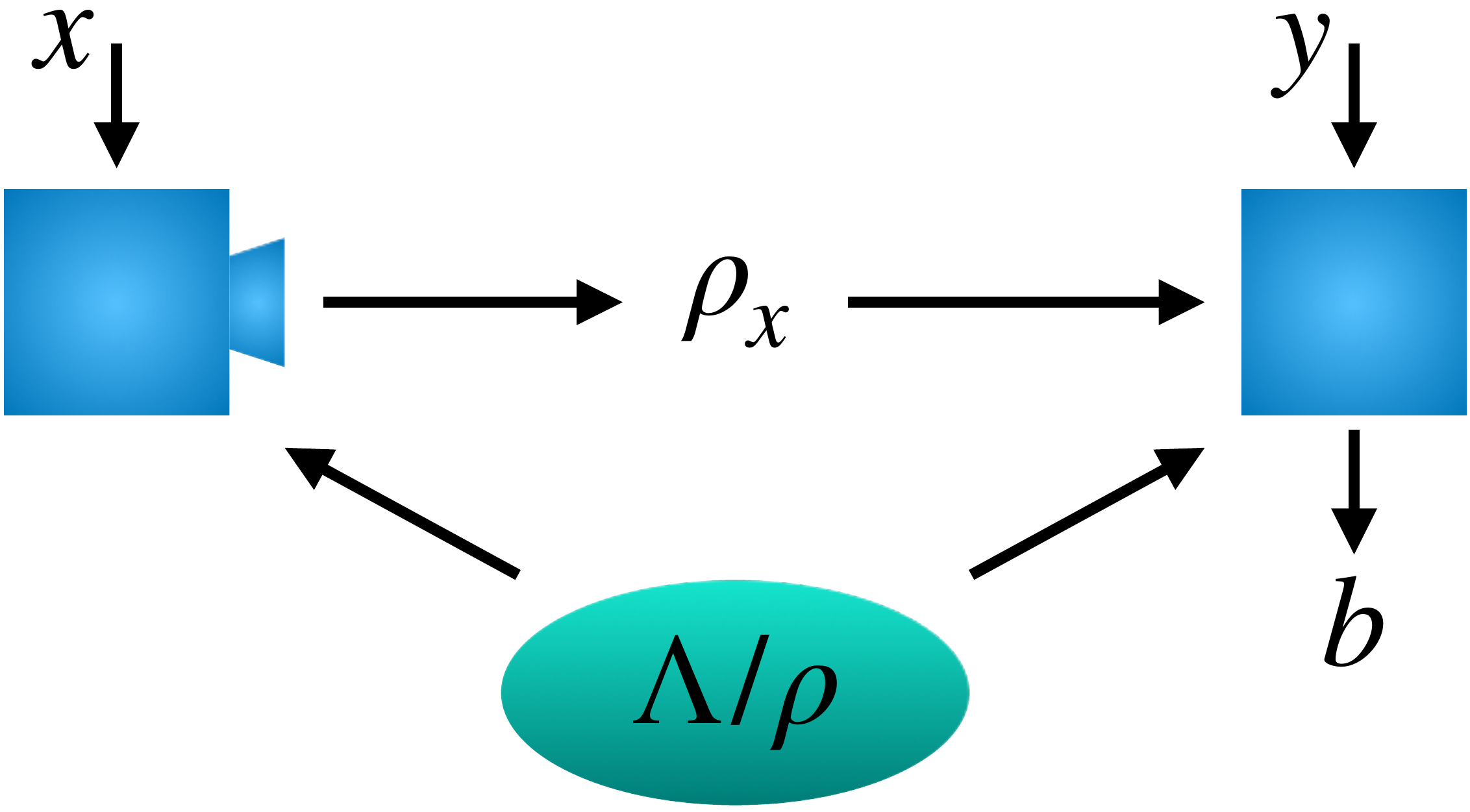}
    \caption{Directed acyclic graph (black box representation) of the prepare and measure scenario where two parties share some correlation, which in principle could be either classical, represented above by the set of variables $\Lambda$, or quantum, represented by a shared state $\rho$. According to some input $x$ Alice prepares a state $\rho_x$ and sends it to Bob, this being the only communication between them, who performs a measurement labeled by some input $y$ obtaining an output $b$.}
    \label{fig: causal structure}
\end{figure}

The prepare and measure scenario consists of an experiment performed between two parties, which we will label Alice and Bob. Alice prepares a system in a state represented by  $x\in\{0,\dots,N-1\}$ and sends it to Bob, who chooses a measurement setting $y\in\{0,\dots,m-1\}$ and obtains an output $b\in\{0,\dots,k-1\}$ (see Fig. \ref{fig: causal structure}). The whole experiment is described by the conditional probability distribution $p(b|x,y)$.

In a classical description, depending on her input $x$, Alice prepares a message $a\in\{0,\dots,l-1\}$, where $l$ is the size of the alphabet of the message $a$, or the dimension of the system, that is a probabilistic function not only of $x$ but also of $\lambda$, the source of possible preshared correlations between Alice's preparation and Bob's measurement apparatus. Similarly, Bob's measurement outcome will depend on the message $a$ being received, the choice of measurement $y$ and the pre-shared correlations. Thus, if the observed distribution has a classical explanation, it can be written as
\begin{eqnarray}
\label{eq:clas}
p(b|x,y) = \sum_{a\in A}\sum_{\lambda\in\Lambda}p(\lambda)p(a|x,\lambda)p(b|a,y,\lambda).
\end{eqnarray}

In turn, a quantum description will explicitly depend on which resources are made nonclassical. For instance, Alice might be allowed to prepare and send quantum states to Bob, but only share classical correlations with him. In this case, the prepared states are described by the set $\{\rho_x\}_{x=0,\dots,N-1}\subset\mathcal{D}(\mathcal{H})$, where $\mathcal{D}(\mathcal{H})$ represents the set of density operators acting on some Hilbert space $\mathcal{H}$. A set of positive semidefinite operators $\{M_{b}^{(y)}\}_{b=0,\dots,k-1}\subset \mbox{Pos}(\mathcal{H})$ for $y=0,\dots,m-1$, for which $\sum_{b=0}^{k-1}M_b^{(y)}=\Id$ $\forall$ $y$, describes the possible measurements performed by Bob. By Born's rule, the observed distribution is then given by
\begin{eqnarray}
p(b|x,y) = \tr \left(\rho_xM_b^{(y)}\right).
\end{eqnarray}
In particular, notice that the quantum and classical descriptions become equivalent if the prepared states $\rho_x$ form a mutually commuting set.

\begin{figure}
    \centering
    \includegraphics[scale = 0.47]{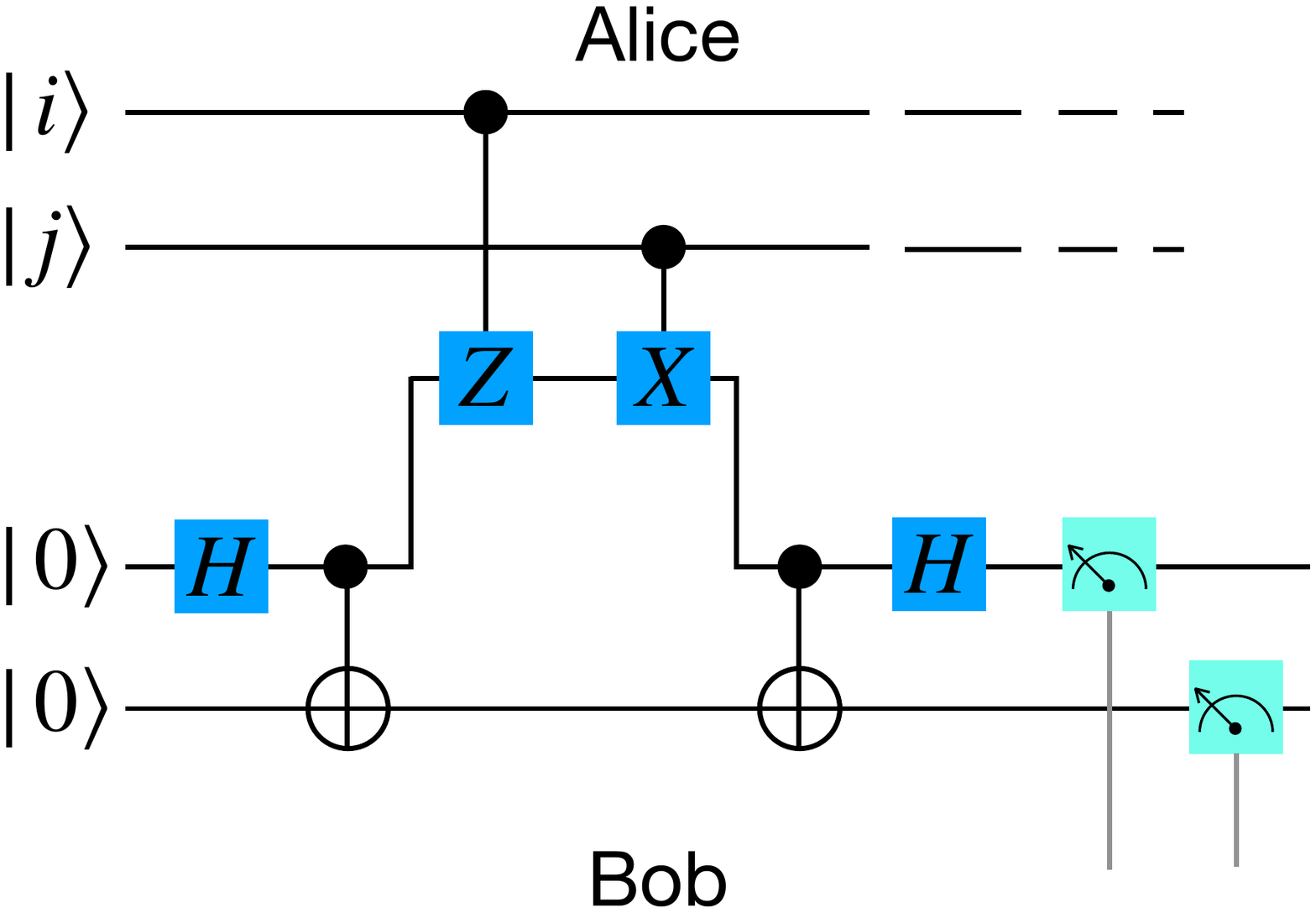}
    \caption{Quantum circuit (device-dependent) representation of the superdense coding. Alice wants to send a two bit message to Bob, represented here by the states $|i\rangle,|j\rangle\in\{|0\rangle,|1\rangle\}$ by sharing an entangled state with Bob (the two bottom qubits in the circuit, the first of which is held by Alice). The goal is achieved by applying $\sigma_z$ conditioned to $|i\rangle$ and $\sigma_x$ conditioned to $|j\rangle$ on the qubit in possession of Alice, which is, then, sent to Bob, who, in turn, retrieves the values of $i$ and $j$ by performing a Bell-state measurement on both qubits.}
    \label{fig:my_label}
\end{figure}

In the most general case, Alice not only prepares and sends quantum states to Bob but might also share entangled states with him. That is precisely the case of the paradigmatic superdense coding protocol \cite{Bennett1992}, where, by sharing entanglement with Bob, Alice can send him 2 dits of information by actually transmitting only one qudit. To illustrate this, suppose Alice wants to send two bits of information to Bob, $(x_0,x_1)\in\left\{ 00,01,10,11 \right\}$. If they share a maximally entangled state $\ket{\Phi^+}=\left( \ket{00}+\ket{11} \right)/\sqrt{2}$, Alice can encode the information to be sent in different local unitaries applied  to the qubit in her possession, for instance  $\left\{ 00,01,10,11 \right\} \rightarrow \left\{ \Id,\sigma_z,\sigma_x,\sigma_z\sigma_x \right\}$, where $\sigma_i$ are the Pauli matrices. Thus, after Alice's local operation, the entangled state shared between them corresponds to the orthonormal Bell basis $\left\{ \ket{\Phi^+},\ket{\Phi^-},\ket{\Psi^+},\ket{\Psi^-}  \right\}$ (depending on which bits Alice wants to send) and that can be discriminated if Alice sends her qubit to Bob and Bob measures both qubits in his possession in the Bell basis. Notice that in this formulation, however, for the superdense coding protocol to work, not only does Alice have to know her state preparations but also Bob has to be sure he is measuring in the Bell basis (see Fig. \ref{fig:my_label}). That is, both the preparation and measurement devices have to be under full control of the parties and be well characterized. In this standard form, the superdense coding protocol is device dependent.

The first hint for the possibility of a semi-DI formulation of the superdense coding is given by the fact that it can be understood as a particular instance of a PAM scenario: one where both the states being communicated as well as the correlations shared between the preparation and measurement devices are quantum. The scenario can be described as follows. Consider a set of states $\{\rho_x\}_{x=0,\dots,N-1}\subset\mathcal{D}(\mathcal{H}_A\otimes\mathcal{H}_B)$ and a set of positive semidefinite operators $\{M_{b}^{(y)}\}_{b=0,\dots,k-1}\subset \mbox{Pos}(\mathcal{H}_A\otimes\mathcal{H}_B)$ for $y=0,\dots,m-1$, for which $\sum_{b=0}^{k-1}M_b^{(y)}=\Id$ $\forall$ $y$. Notice that $\mathcal{H}_A$ and $\mathcal{H}_B$ represent the Hilbert spaces of the systems held by Alice and Bob respectively, such that $\dim(\mathcal{H}_A) = d_A$ and $\dim(\mathcal{H}_B) = d_B$. Thus, the observed probability distribution obtained in the PAM scenario describing superdense coding is given by
\begin{eqnarray}
p(b|x,y) = \tr \left(\rho_xM_b^{(y)}\right),
\end{eqnarray}
where, necessarily,
\begin{eqnarray}
\label{eq: new condition}
\tr_A(\rho_x)=\tr_A(\rho_{x'})\;\;\;\;\forall\; x,x'.
\end{eqnarray}
The condition above is crucial, since it subsumes the idea that Alice's operations (encoding the message she wishes to send) are local and thus cannot affect the marginal quantum state of Bob. Notice that if Alice aims to send two dits of information to Bob, this will correspond to $\vert x\vert=N= d^2$ preparations of Alice. Also notice that in the standard superdense coding, $\vert y \vert =m=1$, that is, Bob always measures the same observable. In this case, assuming that all preparations (the possible values of the dits Alice wishes to send) are equiprobable, we can define a measure of the superdense coding success as
\begin{eqnarray}
\label{eq: single measurement set}
p_{suc} = \frac{1}{N}\sum_{x=0}^{N-1} P(b=x|x).
\end{eqnarray}
We highlight that this is a device-independent measure of success, since it only depends on observational data and does not assume anything about Alice's preparations or Bob's measurements.

\subsection{The Schmidt number}A concept that will play a fundamental role in our results is that of entanglement and its detection via the Schmidt number \cite{Terhal2000}. Any pure bipartite state $|\Psi\rangle\in\mathcal{H}_A\otimes\mathcal{H}_B$ can be represented as
\begin{eqnarray}
\label{eq: Schmidt decomposition}
|\Psi\rangle = \sum_{j=0}^{r-1}\eta_{j}|\psi_j\rangle\otimes|\phi_j\rangle,
\end{eqnarray}
in which $\eta_{j}$ are real positive numbers which are ordered in a way that $\eta_0\geq\eta_1\geq\dots\geq\eta_{r-1}$. The Schmidt rank $r$ of $|\Psi\rangle$ is such that  $1\leq r\leq \min(d_A,d_B)$.
Importantly, the notion of Schmidt rank can be generalized for mixed states via the concept of Schmidt number \cite{Terhal2000}. The Schmidt number $s$ of a mixed state $\rho = \sum_j p_j\ket{\Psi_j}\bra{\Psi_j}$ is defined via an optimization over all possible pure decompositions $\left\{ \ket{\Psi_j} \right\}$ of $\rho$. The Schmidt number $s$ is the smallest possible highest Schmidt rank of the pure states $\ket{\Psi_j}$, that is,
$s=\min_{\left\{ \ket{\Psi_j} \right\}}\max_{r(\ket{\Psi_j})}$. Clearly, for pure states, the Schmidt number coincides with the Schmidt rank. Furthermore, this concept allows a natural classification of the set of bipartite quantum states given by the set $S_k\subseteq \mathcal{D}(\mathcal{H}_A\otimes\mathcal{H}_B)$ composed by all the states with Schmidt number less than or equal to $k$. Those sets are trivially convex and their extremal points are given by pure states, being also clear that $S_1\subset S_2\subset\dots\subset S_{\min(d_a,d_b)}$.

\section{Semi-DI entanglement certification in the superdense coding}\label{sec:Semi-DI} Interestingly, if the preparation and measurement devices are allowed to share only classical correlations, the probability of success \eqref{eq: single measurement set} is the same irrespective of whether Alice sends classical or quantum states to Bob \cite{Brunner2013}. In both cases the probability of success is bounded as $p_{suc} \leq \frac{d_A}{N}$, where $d_A$ is the dimension of the classical or quantum system Alice sends to Bob. This can be seen as a consequence of Holevo's bound \cite{Holevo1973, nielsen2001} that limits the amount of information that may be retrieved in such a scenario, implying that quantum messages cannot transmit more information than their classical counterparts.

However, as shown by the superdense coding protocol, that is no longer the case if an entangled state is shared between the parties. Our first result, for which a detailed proof is given in the Appendix \ref{App: Scenario of superdense coding}, is a formal and quantitative proof of that claim. It shows that the optimal probability of success depends not only on the dimension of the quantum system communicated from Alice to Bob, but also on the amount of entanglement of the quantum state shared between them, as quantified by the Schmidt number.

\begin{result}\label{res:result1}
In a prepare and measure scenario with $N$ preparations and a single measurement with $N$ outcomes, the superdense coding probability of success \eqref{eq: single measurement set} is limited as
\begin{eqnarray}
\label{eq: general single measurement set}
p_{suc}\leq\min\left(\frac{d_As}{N},1\right),
\end{eqnarray}
where $d_A$ is the Hilbert-space dimension of the quantum system sent from Alice to Bob and $s$ is the Schmidt number of the quantum state shared between Alice and Bob. For $N=d_A K$, with $K \geq s$, the bound is tight.
\end{result}

In particular, we notice that for $s=1$, that is, only classical correlations are shared between Alice and Bob, we recover the usual Holevo bound $p_{suc} \leq d_A/N$.

A direct application of the result above is in the context of semi-DI certification of entanglement. The semi-DI comes from the fact that we have to assume the Hilbert-space dimension $\mathcal{H}_A$ to be at most $d_A$. As discussed before, unless one limits the amount of communication sent by Alice, the problem becomes trivial. Since for any separable state $p_{suc}\leq\frac{d_A}{N}$, any probability of success violating this bound is then an unambiguous proof that the shared state must be entangled. We will consider a range of examples of shared states $\rho\in\mathcal{D}(\mathcal{H}_A\otimes\mathcal{H}_B)$, $\dim(\mathcal{H}_A) = d_A$ and $\dim(\mathcal{H}_B) = d_B$, for which a set of $N=d_A^2$ preparations is enough to violate the classical bound that can be rewritten as $p_{suc}\leq\frac{1}{d_A}$.

If $\rho$ is the state under test, we can always define the set of states being prepared by Alice $\{\rho_{x}\}_{x=0,\dots,N -1}$ as
\begin{eqnarray}
\rho_{x} = (\Lambda_{x}\otimes\Id)[\rho],
\end{eqnarray}
in which $\Lambda_{x}$ is a local channel, $\Lambda_{x}:\mathcal{D}(\mathcal{H}_A)\mapsto\mathcal{D}(\mathcal{H}_A)$, for all $x$. Since $\Lambda_{x}$ is a local channel, all the states in $\{\rho_x\}_{x=0,\dots,N-1}$ have a Schmidt number less than or equal to that of $\rho$, so witnessing that $\{\rho_x\}_{x=0,\dots,N-1}$ is not contained in $S_s$ is sufficient to witness that $\rho\not\in S_s$.

Our next result, proven in the Appendix \ref{App: Scenario of superdense coding}, states that every pure bipartite entangled state allows for a quantum enhancement in the superdense coding protocol.

\begin{result}
For $N=d_A^2$, the probability of success in the superdense coding that can be achieved with a bipartite pure entangled state $|\Psi\rangle = \sum_{j=0}^{s-1}\eta_j|j\rangle\otimes|j\rangle$ is lower bounded as
\begin{eqnarray}
\nonumber
p_{suc} & \geq & \frac{1 + \Gamma}{d_A}
\end{eqnarray}
with $\Gamma\equiv \sum_{j\neq k}\eta_j\eta_k>0$ and which violates the classical bound for any nonseparable state ($s >1$).
\end{result}
In particular, notice that for maximally entangled states of dimension $d_A$ we have coefficients $\eta_i=1/\sqrt{d_A}$ and then $\Gamma=(d_A-1)$ implying that $p_{suc}=1$.

Our next result, the proof of which is given in the Appendix \ref{App: Scenario of superdense coding}, connects an important entanglement quantifier with the probability of success in the semi-DI superdense coding. More precisely, we consider the maximal singlet fraction \cite{Horodecki1999} of a general bipartite quantum state $\rho$, given by
\begin{equation}
\zeta(\rho)= \max_{\Phi} \bra{\Phi} \rho \ket{\Phi},     
\end{equation}
where, for some unitary operators $U_1$ and $U_2$, $\ket{\Phi}=(U_1 \otimes U_2)\ket{\Phi^+_{d_A}}$ with $\ket{\Phi^+_{d_A}}=(1/\sqrt{d_A})\sum_{i=0}^{d_A-1} \ket{ii}$  being the maximally entangled state of dimension $d_A$.

\begin{figure}[h]
    \centering
    \includegraphics[width=0.8\columnwidth]{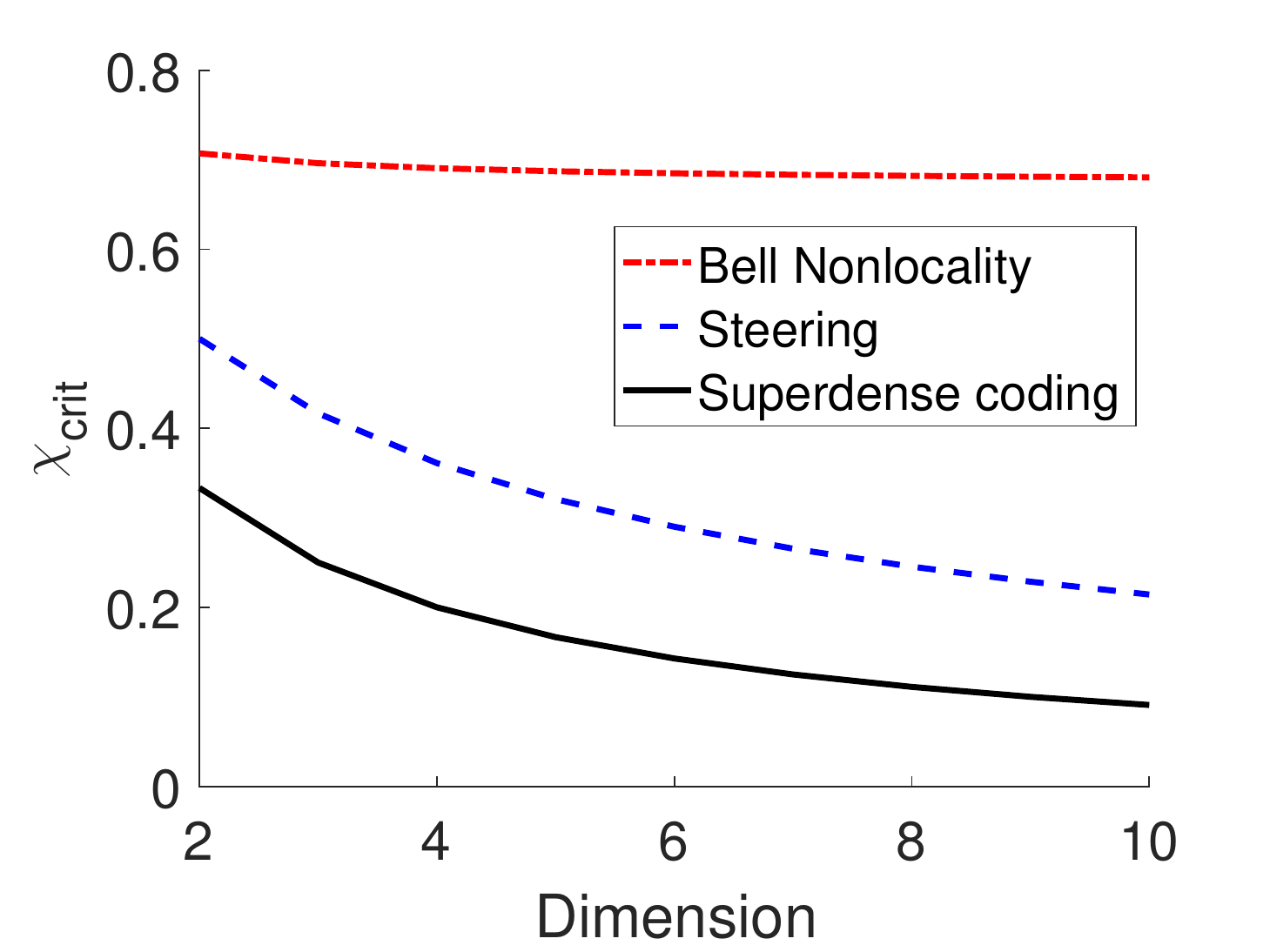}
    \caption{Upper bounds for critical visibilities for entanglement detection in the isotropic state [Eq.\ \eqref{eq:IsotropicState}] with three different methods: Bell nonlocality (red, dot-dashed curve), quantum steering (blue, dashed curve) and superdense coding (black, solid curve). Solely with assumptions on the dimensionality of the distributed system, superdense coding enables entanglement detection for all entangled isotropic states, thus, with lower visibilities as compared with quantum steering \cite{Wiseman2007} and Bell nonlocality. Values for Bell nonlocality were obtained from the violation of the Collins-Gisin-Linden-Massar-Popescu inequality \cite{CGLMP}. Better estimates are known for $d=2$ and $d \rightarrow \infty$ \cite{Cavalcanti2011}, which reduce $\chi^B_{crit}$ to $0.67$ and $0.5$, respectively, but are still greater than the value provided by the superdense coding method. }
    \label{fig:IsotropicComparison}
\end{figure}

\begin{result}
The best probability of success in the superdense coding method provided by a shared bipartite state $\rho$ is lower bounded as
\begin{eqnarray}
\label{eq:pchi}
p_{suc} & \geq & \zeta(\rho),
\end{eqnarray}
in which $\zeta(\rho)$ is the maximal singlet fraction of $\rho$, and $d_A$ is the dimension of the quantum state sent from Alice to Bob.
\end{result}

As a particular case we can consider the family of isotropic states, a usual benchmark for the utility of a nonclassicality witness \cite{Cavalcanti2011}. These states are given by $\rho_{\chi}\in\mathcal{D}(\mathcal{H}_A\otimes\mathcal{H}_B)$, for $\dim (\mathcal{H}_A)= \dim(\mathcal{H}_B) = d_A$ with
\begin{eqnarray}
\rho_{\chi} = \frac{(1-\chi)}{d_A^2}\Id + \chi \ket{\Phi^+_{d_A}}\bra{\Phi^+_{d_A}}.
\label{eq:IsotropicState}
\end{eqnarray}
As can be seen, the singlet fraction is given by $\zeta(\rho_{\chi}) = \chi + (1-\chi)/d_{A}^{2}$. In particular, the critical visibility below which the isotropic state becomes separable is given by
\begin{eqnarray}
\chi_{crit} & = & \frac{1}{d_A+1},
\end{eqnarray}
which coincides with the critical $\chi$ below which the probability  of success \eqref{eq:pchi} becomes classical. Strikingly, our semi-DI witness detects the nonclassicality of any entangled isotropic state.

It has been recently proved that a state $\rho \in\mathcal{D}(\mathcal{H}_A\otimes\mathcal{H}_B)$, for $\dim (\mathcal{H}_A)= \dim(\mathcal{H}_B) = d_A$ is a \emph{faithful entangled state}, \textit{i. e.}, its entanglement can be detected by a fidelity-based test, if and only if, $\zeta(\rho) > 1/d_{A}$ \cite{Guehne2021}. Combined with Result 3, this implies that our semi-DI witness can, in fact, detect the non-classicality of \emph{all} faithful entangled states, a set of which the isotropic states previously mentioned are particular examples.

As a comparison we can consider the paradigmatic Bell \cite{brunner2014bell} and steering tests \cite{Wiseman2007}, both involving shared entangled states and measurement devices for both Alice and Bob. A Bell test is fully DI and for this reason leads to higher constraints over $\chi$. In turn, the steering scenario is semi-DI, because tomography on Bob's state is required, thus implying not only that the dimension of the state has to be known but also that one has to trust the measurement device. In this sense, the semi-DI requirements in the superdense coding protocol are milder as compared to the steering, since the former only require an assumption on the state dimension. 
For $d=2$, the best known Bell test \cite{Cavalcanti2011}  requires $\chi^B_{crit}=0.64$. For steering, one gets \cite{Wiseman2007} $\chi^S_{crit}=0.5$, while for the superdense coding we get $\chi^{SD}_{crit}=0.33$. In turn, making $d \rightarrow \infty$ the best known Bell test implies $\chi^B_{crit}=0.5$, while for a steering test $\chi^S_{crit}=(H_{d_A}-1)/(d-1)$, where $H_n = \sum_{i=1}^n 1/i$ is the $n$th harmonic number, implying that for any dimension there will be a gap between the steering and the superdense coding or entanglement tests. See Figure \ref{fig:IsotropicComparison} for more details.

\section{Self-testing maximally entangled states}

An important application of the DI framework is the possibility to infer properties of the shared quantum state without the need of knowing precisely the measurement apparatus, a feature known as self-testing \cite{mayers2003self,vsupic2020self}. As we show next, under the assumption of the dimension $d_A$ of the shared bipartite state, the PAM scenario can be employed to self-test maximally entangled states.

\begin{result} For $N=d_A^2$, the saturation of inequality \eqref{eq: general single measurement set} for $s=d_A$ self-tests, up to a local unitary, the presence of a bipartite maximally entangled state. \end{result}

It is worth highlighting that self-testing in the superdense coding is likewise the one in a Bell scenario and differs from usual self-testing results in PAM scenario \cite{tavakoli2018self,miklin2020universal}. Typically, the PAM scenario without shared entanglement can self-test a set of prepared states. Here, in contrast, we are self-testing the shared quantum state and not the preparations.

Another curious feature of this self-testing process relies on a strong dependence on the hypothesized causal structure. Self-testing in superdense coding only certifies that Alice and Bob share a maximally entangled state with someone else, but not necessarily with each other. For instance, Alice and Bob might share a maximally entangled state with an eavesdropper and still saturate the superdense coding witness \eqref{eq: general single measurement set}. This is an unusual feature, for instance, when compared with self-testing in Bell scenarios that are robust to the insertion of an extra part, a crucial property in applications such as quantum key distribution. In the case of the prepare and measure scenario, entanglement might be used to break existing semi-DI quantum key distribution protocols \cite{pawlowski-qkd-2011}.

This shows that even though our witness is semi-DI (as it only assumes the dimension of the state but no other information from the preparation and measurement devices), in principle it is not robust against an external malicious part. As pointed out above, in the superdense coding an eavesdropper can retrieve the information being sent from Alice to Bob without being detected. The source of cryptographic insecurity comes from the fact that the measurement device of Bob has a single input, thus allowing the eavesdropper (sharing entanglement with Bob) to retrieve the information without being detected.

To avoid that, at least in a device dependent framework, one possibility is to adapt the BB84 protocol \cite{bennett2014quantum}. Say that Alice randomly decided whether or not to apply a Hadamard gate $H$ --- such that $H\ket{0}=(1/\sqrt{2})(\ket{0}+\ket{1})$ and $H\ket{1}=(1/\sqrt{2})(\ket{0}-\ket{1})$ --- to the qubit she sends to Bob. Without knowing if Alice applied or not the Hadamard gate, the eavesdropper will unavoidably make detectable mistakes.

For instance, if Alice wanted to send the classical message $00$ and did not apply the Hadamard to her qubit, in the absence of an eavesdropper the state Bob would receive is $(1/\sqrt{2})(\ket{00})+\ket{11}$. The eavesdropper, however, does not know whether the Hadamard was applied or not. If he randomly decides to apply the Hadamard gate to the qubit he intercepts (even though Alice has not done it) he will then with probability half wrongly conclude that the message being sent by Alice was $10$. The eavesdropper will not only resend the wrong information to Bob but also with probability one-half he will choose the wrong encoding. If, similarly to what happens in the BB84 protocol, Alice and Bob use some rounds to publicly compare their encoded and decoded messages, they will unavoidably detect the presence of the eavesdropper.

In summary, combining the BB84 the superdense coding protocol  makes the latter robust in a cryptographic sense \cite{PhysRevA.69.032310}. Notice, however, that in this case the protocol becomes device dependent (we have explicitly used the quantum description of a Hadamard gate). A possibility to achieve a semi-DI cryptographic formulation would be to use a witness such that the measurement device takes more than just one possible input. We derive an example of such witness below but leave open the possibility of whether it allows one to secure the flow of quantum information from an eavesdropper.

\section{Optimizing the probability of success} Although useful lower bounds for the best probability of success can be found analytically for specific states by taking advantage of their special structures, in a more general case, for a generic shared state, finding good guesses for preparations and measurements might be a cumbersome task. An interesting alternative is to find such lower bounds numerically. Given some state shared between Alice and Bob, in order to achieve the optimal probability of success one has to optimize over all possible preparations of Alice and the possible measurements of Bob. As we show next, this optimization can be performed via semidefinite programming \cite{boyd2004}. In particular, if the preparations of Alice are fixed, optimization over Bob's measurement is given by the following program:
\begin{align}
\textrm{Given} &\quad \rho_{x} = (\Lambda_{x}\otimes\Id)[\rho], \nonumber \\
\maximize_{M_x} &\quad \sum_x\tr[\rho_x\,M_x], \nonumber \\
\textrm{subject to} &\quad M_x \geq 0, \nonumber \\
&\quad \sum_x M_x = \mathbbm{1}.
\end{align}

\begin{figure}[t!]
    \centering
    \includegraphics[width=\columnwidth]{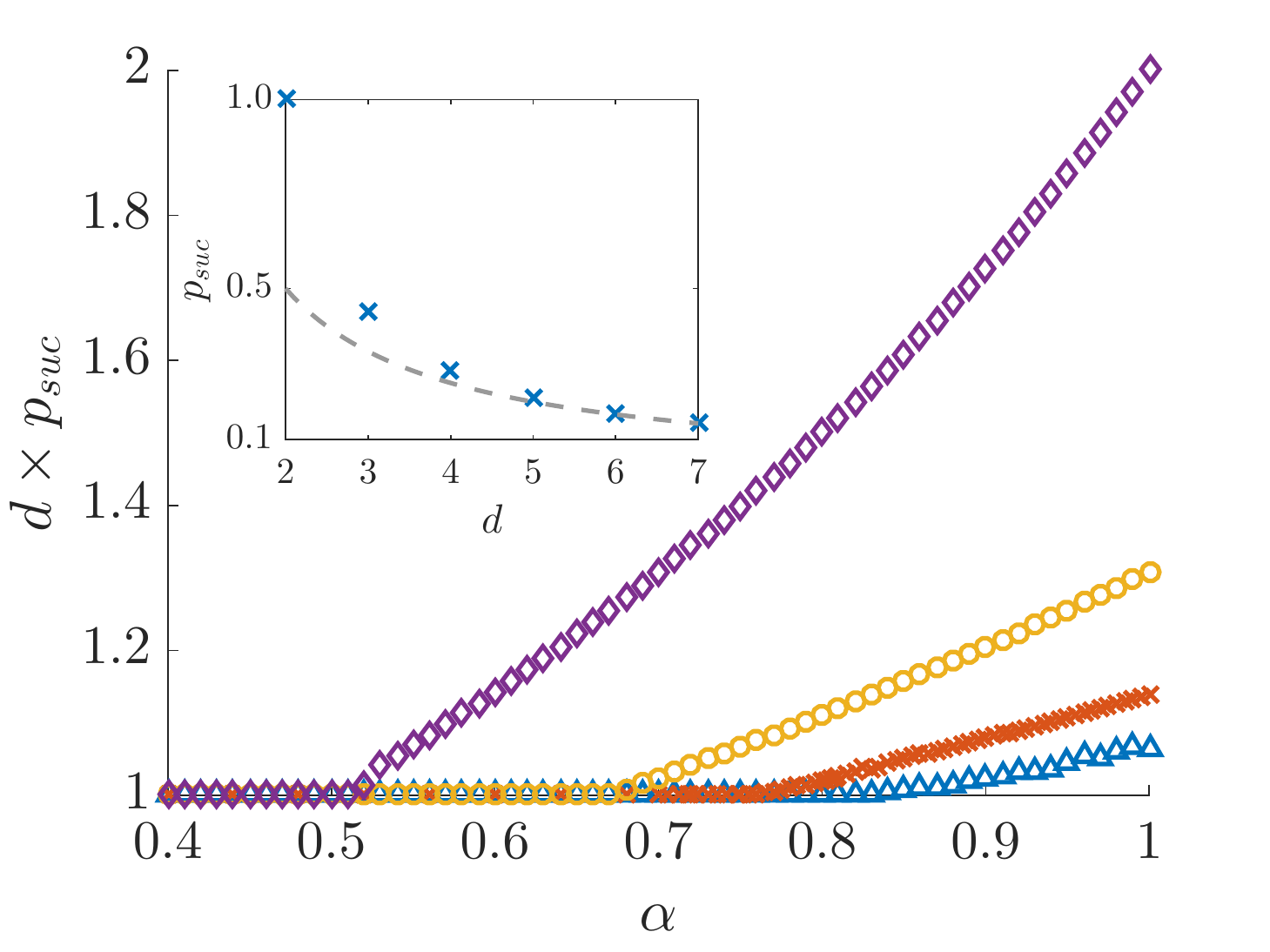}
    \caption{Lower bounds on success probabilities for super dense coding, computed via the alternated optimization method for Werner states. The curves correspond to different local dimensions $d$, ranging from 2 to 5, from top to bottom (purple rhombuses correspond to $d=2$, yellow circles to $d=3$, red crosses to $d=4$, and blue triangles to $d=5$). Values are rescaled by $d$, so that the bounds for separable states for all cases coincide in value $1$. In every case computed, detection of entanglement occurs after $\alpha \approx (d-1)/d$. Inset:  Comparison between $p_{suc}$ for $\alpha=1$ (blue crosses) and the classical bound $1/d$ for dimensions $2,...,7$ (dashed curve).}
    \label{fig:UN_Werner4}
\end{figure}

In turn, fixing the measurements of Bob allows for optimization over possible preparations in terms of an SDP by using the Choi-Jamiolkowski representation of the different channels \cite{Choi, Jamiolkowski} as
\begin{align}
\maximize_{L_x} &\quad \sum_x\tr[(L_x \otimes \mathbb{1}_B) (\rho^{T_A} \otimes \mathbbm{1}_{A^\prime}) \left( \mathbbm{1}_{A} \otimes M_x \right)], \nonumber \\
\textrm{subject to} &\quad L_x \geq 0, \nonumber \\
&\quad \tr_{A^\prime}[L_x] = \mathbbm{1}_{A},
\end{align}
where $L_x$ are the operators that act on $\mathcal{H}_{A} \otimes \mathcal{H}_{B}$ and correspond to each preparation $\Lambda_x$, $\rho^{T_{A}}$ is the partial transposition of the state $\rho$ shared between Alice and Bob, and the constraints ensure that the resulting maps are completely positive and trace preserving.

By alternating between preparation optimization and measurement optimization, starting with a random measurement, we can obtain a lower bound on the optimal $p_{suc}$ for any given state $\rho$. 

As an application of this method, we consider the Werner states \cite{Werner1989}, described by
\begin{equation}
\rho_W(\alpha) = \frac{\mathbbm{1}_{d^2} - \alpha S}{d^2 - \alpha d},
\end{equation}
where $d$ is the local dimension of each subsystem, $S$ is the swap operator $\sum_{i,j=1}^d |ij\rangle\langle ji|$, and $\alpha$ is a parameter in the range $[-1,\,1]$. Using $N=d^2$ preparations and outcomes for Bob's measurement and applying the method to $\rho_W(\alpha)$ for $d=2,\ldots,5$, we find that $p_{suc}$ saturates the bound $1/d$ [Eq.\ \eqref{eq: general single measurement set}] for all values of $\alpha$ below approximately $(d-1)/d$, and violates the bound for all values above. This is shown in Fig. \ref{fig:UN_Werner4}, where $p_{suc}$ is plotted for $\alpha \geq 0.4$. Remarkably, the threshold at $(d-1)/d$ is strictly lower than the threshold for establishing quantum steering \cite{Wiseman2007}, given by $d/(d+1)$. That is, the semi-DI test provided by the superdense coding once more beats steering tests. In the inset of  Fig. \ref{fig:UN_Werner4} we show the values computed for $p_{suc}$ when $\alpha=1$ as a function of the dimension, for dimensions $d=2,\ldots,7$. As can be seen, the gap $p_{suc} - 1/d$ decreases quickly with the dimension, indicating that Werner states of larger dimension provide smaller quantum enhancements in the superdense coding.

\section{prepare and measure scenario with more than one measurement setting}

As discussed above, the fact that in the superdense coding the measurement device of Bob only has one input opens the way to attacks of an external malicious part. An adaptation of the BB84 protocol is enough to guarantee the security of the superdense coding, however, in a device dependent manner. Motivated by that we provide below another prepare and measure test that witnesses the entanglement of the shared quantum state but, in this case, relying on several different measurements of Bob. Whether this witness or a variation of it can be combined with the superdense coding to guarantee its cryptographic security is an interesting problem that we leave open for future research.

Consider the prepare and measure framework featuring $N$ preparations and $N(N-1)/2$ dichotomic measurement settings. This has been analyzed in Ref. \cite{Brunner2013} under the hypothesis that the shared correlations are classical. Here we drop this hypothesis and show that new bounds must be considered when the participants share a quantum state. Remarkably, in this case, as in the superdense coding scenario, we observe that the Schmidt number of the shared state plays a central role in defining the bound.

\begin{result}
In a prepare and measure scenario with $N$ preparations $x$ and $N(N-1)/2$ dichotomic measurements $(y_1,y_2)$, for $y_1>y_2$, where $y_1,y_2\in\{0,\dots,N-1\}$, the set of probability distributions  is bounded by the inequality:
\begin{eqnarray}
\label{eq: general N(N-1)/2 measurements set}
V_N & \leq & \frac{N^2}{2}\left(1 - \frac{1}{\min (d_As, N)}\right),
\end{eqnarray}
where $d_A$ is the Hilbert-space dimension of the quantum system sent from Alice to Bob, $s$ is the Schmidt number of the quantum state shared between Alice and Bob, and
\begin{eqnarray}
V_N & = & \sum_{x>x'} \left|P(1|x,(x,x')) - P(1|x',(x,x')\right|^2.
\end{eqnarray}
Furthermore, if $s=d_A$ and $N< d_A^2$ or $N=cd_A^2$, for integer $c$, expression \eqref{eq: general N(N-1)/2 measurements set} is tight.
\end{result}

A detailed proof of the results is provided in the Appendix \ref{App: N preparations and N(N-1)/2}.

\section{Discussion}

The ability to certify entanglement is a crucial benchmark in quantum information processing. A standard tool for that is entanglement witnesses \cite{Horodecki2009}, experimentally observable quantities allowing one to distinguish between separable and entangled states. However, and in spite of its wide applicability, entanglement witnesses are fully device dependent. Unless one has perfect characterization of measurement devices, one might incur false positive results \cite{rosset2012}.

The best one can hope for is a fully device-independent certification of entanglement, such as that provided by the violation of Bell inequalities \cite{brunner2014bell}. The problem, however, is the fact such violations are still an experimental challenge and furthermore are unable to witness the nonclassicality of a wide range of entangled states \cite{Werner1989,Wiseman2007}. A promising approach to achieve a compromise between our ability to witness entanglement with fewer assumptions as possible and at the same time to achieve experimental feasibility is that offered by semi-device-independent protocols. In quantum steering \cite{Wiseman2007}, for instance, one can detect a larger set of entangled states at the cost, however, of being able to perform quantum tomography on some parts of the entangled system.

Here we show that a paradigmatic protocol in quantum information science, the superdense coding protocol \cite{Bennett1992}, offers a platform for entanglement certification. As we show, superdense coding can be seen as a particular case of a prepare and measure scenario \cite{gallego-pam-2010}, one where both the communication and the shared correlations are allowed to have a quantum nature. Within this context, we provide a semi-device-independent witness---requiring only an assumption on the Hilbert-space dimension of the quantum state---that is upper bounded by the Schmidt number of the shared quantum state. This witness not only has a clear operational meaning---the probability of success of the superdense coding protocol---but also can be connected to an important entanglement quantifier, the so-called singlet fraction \cite{Horodecki1999}, implying in particular that any pure bipartite entangled state offers a semi-DI advantage in the superdense coding protocol. Furthermore, our approach provides a significant advantage in comparison with steering \cite{Wiseman2007}, the standard semi-DI test in the literature. As opposed to a steering test, our witness not only does not require quantum state tomography but also can witness the nonclassicality of any entangled isotropic state, an important family of mixed entangled states used as a benchmark in DI and semi-DI certification of entanglement. Nicely, our witness can also be used to self-test maximally entangled states of any dimension. Finally, we provide a semidefinite program formulation allowing one to obtain, for any shared quantum state, lower bounds for the best probability of success that can be obtained in the execution of superdense coding.

In the scenario where no shared quantum correlations are allowed, the prepare and measure scenario has been employed in a variety of quantum information tasks \cite{bowles2015testing,wang2019characterising,tavakoli2018self,miklin2020universal,li2012semi,pawlowski-qkd-2011,passaro-randomness-2015}. Thus, an interesting question is whether the fully quantum version of the PAM scenario we consider here can also lead to relevant practical applications. For instance, we have shown that in its standard form, the dense coding is not cryptographically secure, as a malicious part could retrieve the information being sent without being detected. As discussed, the source of insecurity comes from the fact that the measurement device has a single input. This has motivated us to also derive a witness with several measurement inputs. Perhaps a combination of both witnesses could provide the desired cryptographic security. Another possibility is to investigate whether the PAM scenario with quantum correlations can also be employed to detect the dimension of physical systems. So far, dimension witnesses \cite{gallego-pam-2010,chaves2015device,van2017semi,tavakoli2020characterising,tavakoli2020informationally,poderini2020criteria} make the strong assumption that only classical correlations are allowed between the preparation and measurement devices, an assumption that if not fulfilled ruins the current applications of such witnesses \cite{pawlowski-qkd-2011,passaro-randomness-2015,li2012semi}.
We believe our results might trigger further developments in this direction.

\vspace{1cm}

\noindent\textbf{Note added}: After publication of this work it came to our attention the results of Ref. \cite{Tavakoli2018} concerning multipartite entanglement certification, which in principle describes a scenario more restrictive than the one introduced in this manuscript, since there it is assumed that no correlations are allowed between the set of senders $\{A_1,\dots,A_n\}$ and the receiver $B$, and each subsystem's dimension is assumed to be upper bounded. However, as an intermediate step in their proof they also obtain our result \ref{res:result1} for $s=1$. There, results connecting the probability of success with the singlet fraction are also obtained (though considering GHZ states).

Also after publication of this work we became aware of the results of Ref.~\cite{Guehne2021}, which imply that the semi-DI witness we introduce are, in fact, able to detect the non-classicality of all faithful entangled states. We added a paragraph explaining this connection in Section \ref{sec:Semi-DI}.

\begin{acknowledgments}
We acknowledge the John Templeton Foundation via Q-CAUSAL Grant No. 61084, the Serrapilheira Institute (Grant No. Serra-1708-15763), the Brazilian National Council for Scientific and Technological Development via the National Institute for Science and Technology on Quantum Information, Grants No. 307172/2017-1 and No. 406574/2018-9, the Brazilian agencies MCTIC and MEC, the S\~{a}o Paulo Research Foundation FAPESP (Grant No. 2018/07258-7) and FAEPEX/UNICAMP (Grant No. 3044/19).
\end{acknowledgments}

\appendix
\setcounter{theorem}{0}

\section{Scenario of superdense coding}
\label{App: Scenario of superdense coding}

In this appendix we provide a detailed proof of the results introduced in the main paper, each of which is restated below for convenience.

\begin{result}\label{res:app_result1}
In a prepare and measure scenario with $N$ preparations and a single measurement with $N$ outcomes, the superdense coding probability of success \eqref{eq: single measurement set} is limited as
\begin{eqnarray}
\label{eq:app general single measurement set}
p_{suc}\leq\min\left(\frac{d_As}{N},1\right),
\end{eqnarray}
where $d_A$ is the Hilbert-space dimension of the quantum system sent from Alice to Bob and $s$ is the Schmidt number of the quantum state shared between Alice and Bob. For $N=d_A K$, with $K \geq s$, the bound is tight.
\end{result}

\begin{proof}
First, let us notice that $p_{suc}$ is a linear function defined in $\mathcal{D}(\mathcal{H}_A\otimes\mathcal{H}_B)$. Since $S_s\subseteq \mathcal{D}(\mathcal{H}_A\otimes\mathcal{H}_B)$ is convex for all possible values of $s$, it must hold that $p_{suc}$ is a convex function defined in $S_s$ for all $s$. 

We are interested in setting an upper bound on the value of $p_{suc}$ for a set $\{\rho_x\}_{x=0,\dots,N-1}\in S_s$ for some fixed $s$. Since $p_{suc}$ is a convex function in $S_s$, its maximum value must happen for extremal points in $S_s$, which are pure states. Thus, we focus on the case $\{|\Psi_x\rangle\langle\Psi_x|\}_{x=0,\dots,N-1}\in S_s$.

Given that Alice's preparations cannot affect Bob's side [Eq.\ \eqref{eq: new condition}] and using the Schmidt decomposition of each $|\Psi_x\rangle$ [Eq.\ \eqref{eq: Schmidt decomposition}], we obtain
\begin{eqnarray}
|\Psi_x\rangle\langle\Psi_x| = \sum_{j,k=0}^{s-1}\eta_j^{(x)}\eta_k^{(x)}|\psi_j^{(x)}\rangle\langle\psi_k^{(x)}|\otimes|\phi_j\rangle\langle\phi_k|,
\end{eqnarray}
for $|\psi_j^{(x)}\rangle\in\mathcal{H}_A$ and $|\phi_j\rangle\in\mathcal{H}_B$.

Let us consider the orthonormal basis of $\mathcal{H}_B$ given by $\{|\phi_j\rangle\}_{j=0,\dots,d_B-1}$, in which for $0\leq j\leq s-1$ the elements $|\phi_j\rangle$ are exactly the same that appear in the Schmidt decomposition of $|\Psi_x\rangle$. Plus, let $\mathcal{H}_{aux}$ be the space generated by the set of orthogonal vectors $\{|\phi_j\rangle\}_{j=0,\dots,s-1}$. Then, we have that $|\Psi_x\rangle\in\mathcal{H}_{effective} = \mathcal{H}_A\otimes\mathcal{H}_{aux}$ for $x\in\{0,\dots,N - 1\}$, and $\dim(\mathcal{H}_{effective}) = d_As$.

Thus, in this case,
\begin{eqnarray}
\nonumber
p_{suc} & = & \frac{1}{N}\sum_{x=0}^{N-1} \tr (|\Psi_x\rangle\langle\Psi_x|M_x)\\
\nonumber
    & = & \frac{1}{N}\sum_{x=0}^{N-1} \tr_{effective} \left(|\Psi_x\rangle\langle\Psi_x|M'_x\right)\\
\nonumber
    & \leq & \frac{1}{N}\sum_{x=0}^{N-1} \tr_{effective} \left(M'_x\right)\\
    & = & \frac{d_As}{N},
\end{eqnarray}
in which $M'_x$ is a positive semidefinite operator acting on $\mathcal{H}_{effective}$ and $\sum_{x=0}^{N-1} M'_x = \Id_{effective}$, where $\Id_{effective}$ is the identity acting on $\mathcal{H}_{effective}$.

To verify that the bound is tight for $N = d_A K$, with $K \geq s$, consider the unitary operators
\begin{equation}
W^{(K)}_{x_1,x_2} = \sum_{j=0}^{d_A-1} e^{2 \pi i j x_2/K} | j \oplus x_1 \rangle \langle j |,
\label{eq:app_ModWeyl}
\end{equation} 
defined for $x_1 \in \{0,\ldots,d_A-1\}$ and $x_2 \in \{0,\ldots,K-1\}$, which coincide with the Weyl operators for $K = d_A$ \cite{Watrous2018}. Assume that Alice's preparations are given by application of the $W^{(K)}_{x_1,x_2}$ on her side of the shared state and that the shared state is maximally entangled with Schmidt rank $s$, i.e. 
\begin{equation}
|\psi\rangle = \sum_{j=0}^{s-1} \frac{1}{\sqrt{s}}|j\rangle|j\rangle.
\end{equation}

Let us define then the resulting states as
\begin{equation}
\label{eq:app_PrepStates}
|\tilde{\Psi}^{(K)}_{x_1,x_2}\rangle \coloneqq \sum_{j=0}^{s-1} \frac{1}{\sqrt{s}}\left(W^{(K)}_{x_1,x_2}|j\rangle\right) |j\rangle.
\end{equation}
It is straightforward to show that, for $K \geq s$, $\,\sum_{x_1,x_2}\,|\tilde{\Psi}^{(K)}_{x_1,x_2}\rangle\langle\tilde{\Psi}^{(K)}_{x_1,x_2}| = (K/s)\,\sum_{j=0}^{s-1} \mathbbm{1}_A \otimes |j\rangle\langle j|$. Assume then that Bob's measurement operators are given by
\begin{equation}
M_{x_1,x_2} = \frac{s}{K}|\tilde{\Psi}^{(K)}_{x_1,x_2}\rangle\langle\tilde{\Psi}^{(K)}_{x_1,x_2}| + \frac{1}{N} \sum_{j=s}^{d_B-1}  \mathbbm{1}_A \otimes |j\rangle\langle j|,
\end{equation}
so that $M_{x_1,x_2}|\tilde{\Psi}^{(K)}_{x_1,x_2}\rangle = (s/K)|\tilde{\Psi}^{(K)}_{x_1,x_2}\rangle$ for all $x_1,\,x_2$. With this prescription, we obtain that
\begin{equation}
p_{suc} = \frac{s}{K}.
\end{equation}
Since $N = d_A K$, we obtain precisely that $p_{suc} = d_A s/N$.
\end{proof}

\begin{result}
For $N=d_A^2$, the probability of success in the superdense coding that can be achieved with a bipartite pure entangled state $|\Psi\rangle = \sum_{j=0}^{s-1}\eta_j|j\rangle\otimes|j\rangle$ is lower bounded as
\begin{eqnarray}
\nonumber
p_{suc} & \geq & \frac{1 + \Gamma}{d_A}
\end{eqnarray}
with $\Gamma\equiv \sum_{j\neq k}\eta_j\eta_k>0$ and which violates the classical bound for any nonseparable state ($s >1$).
\end{result}

\begin{proof}
Let $|\Psi\rangle$ be the state under test and assume its Schmidt rank as $s$, for some $1 \leq s \leq d_A$. Define the set of states
\begin{eqnarray}
\rho^{(x_1,x_2)} = \sum_{j,k=0}^{s-1}\eta_j\eta_k(W_{x_1,x_2}|j\rangle\langle k|W_{x_1,x_2}^{\dagger})\otimes|j\rangle\langle k|,
\end{eqnarray}
where $W_{x_1,x_2}$ are the Weyl operators [$K=d_A$ in Eq.\ \eqref{eq:app_ModWeyl}] and the detection operators as
\begin{eqnarray}
\label{eq: maximally entangled measurement}
\tilde{M}_{b_1,b_2} = |\tilde{\Psi}_{b_1,b_2}\rangle\langle\tilde{\Psi}_{b_1,b_2}|,
\end{eqnarray}
where $|\tilde{\Psi}_{b_1,b_2}\rangle = |\tilde{\Psi}^{(d_A)}_{b_1,b_2}\rangle$ [Eq.\ \eqref{eq:app_PrepStates}]. Then, we obtain
\begin{multline}
\nonumber
\rho^{(x_1,x_2)}M_{x_1,x_2} =\\ \frac{1}{d_A}\sum_{m=0}^{d_A - 1}\sum_{j,k=0}^{s-1}\eta_j\eta_k(W_{x_1,x_2}|j\rangle\langle m|W_{x_1,x_2}^{\dagger})\otimes|j\rangle\langle m |,
\end{multline}
which proves the result.
\end{proof}

\begin{result}
The best probability of success in the superdense coding provided by a shared bipartite state $\rho$ is lower bounded as
\begin{eqnarray}
\label{eq:app_pchi}
p_{suc} & \geq & \zeta(\rho),
\end{eqnarray}
in which $\zeta(\rho)$ is the maximal singlet fraction of $\rho$ and $d_A$ is the dimension of the quantum state sent from Alice to Bob.
\end{result}

\begin{proof}We start by recalling the fact that any state $\rho$ presenting a maximal singlet fraction $\zeta(\rho)$ can be converted via shared randomness and local unitary operations into an isotropic state $\rho_{\chi}$ with same maximal singlet fraction,
\begin{eqnarray}
\rho_{\chi} = \frac{(1-\chi)}{d_A^2}\Id + \chi \ket{\Phi^+_{d_A}}\bra{\Phi^+_{d_A}},
\end{eqnarray}
in which $\chi = \frac{\zeta(\rho) d_A^2 - 1}{d_A^2 - 1}$, via a twirling operation \cite{Quintino2016, Horodecki1999II}. This implies that, without any loss of generality, we can restrict our demonstration to isotropic states only.

Define the states given by
\begin{eqnarray}
\nonumber
\rho_{\chi}^{x_1,x_2} & = & W_{x_1,x_2}\rho_{\chi}W_{x_1,x_2}^{\dagger}\\
\nonumber
& = & \frac{1 - \chi}{d_A^2}\Id + \chi W_{x_1,x_2}|\Phi^+_{d_A}\rangle\langle\Phi^+_{d_A}|W_{x_1,x_2}^{\dagger},
\end{eqnarray}
and the measurement basis defined in equation \eqref{eq: maximally entangled measurement}.

Then, using the fact that $\tr[\rho_{\chi}^{x_1,x_2} M_{x_1,x_2}] = \langle \Phi^+_{d_A}| \rho_{\chi} | \Phi^+_{d_A} \rangle$, we obtain
\begin{eqnarray}
\tr\left[\rho_{\chi}^{x_1,x_2}M_{x_1,x_2}\right] & = & \zeta(\rho_{\chi}),
\end{eqnarray}
and given that $\zeta(\rho_{\chi}) = \zeta(\rho)$, we get
\begin{eqnarray}
\nonumber
p_{suc} = \zeta(\rho).
\end{eqnarray}

Hence, using local operations and shared randomness, it is possible to certify any state with a maximal singlet fraction satisfying
\begin{eqnarray}
\nonumber
\zeta(\rho) & > & \frac{1}{d_A}.
\end{eqnarray}

Remarkably, for isotropic states, this is precisely the condition for nonseparability \cite{Horodecki1999_III}, implying such states are entangled if and only if they can violate our witness.

\end{proof}

\begin{result} The saturation of inequality \eqref{eq: general single measurement set} for $s=d_A$, with $N=d_A^2$ preparations, self-tests the presence of a shared bipartite maximally entangled state up to local unitary operations.  \end{result}

\begin{proof}
First of all, we recall the fact that for a fixed dimension $d_A$ of the Hilbert space of Alice's system, $\mathcal{H}_A$, any set of preparations is contained in an effective Hilbert space of dimension $d_A^2$, $\mathcal{H}_{effective}$.

Given that for reaching such bound we must have $p(b=x|x) = 1$, we can assure that the preparations $\{\rho_x\}_{x=0,\dots,N-1}$ do not overlap, i.e., $\tr(\rho_x\rho_{x'}) = 0$ if $x\neq x'$, otherwise no measurement would perfectly distinguish them.

This, associated with the condition $N=d_A^2$, imposes that the states must also be pure, $\{\rho_x\}_{x=0,\dots,N-1} = \{|\Psi_x\rangle\langle\Psi_x|\}_{x=0,\dots,N-1}$. To get this result, we use the spectral decomposition of $\rho_x$:
\begin{eqnarray}
\rho_x = \sum_{j=0}^{d_A^2-1}\lambda_j^{(x)}|\Psi^{(x)}_j\rangle\langle\Psi^{(x)}_j|,
\end{eqnarray}
in which $\langle\Psi^{(x)}_j|\Psi^{(x)}_k\rangle = \delta_{j,k}$. Then, we obtain
\begin{eqnarray}
\nonumber
\tr(\rho_x\rho_{x'}) & = & \sum_{j,k=0}^{d_A^2-1}\lambda_j^{(x)}\lambda_k^{(x')}\tr(|\Psi^{(x)}_j\rangle\langle\Psi^{(x)}_j|\Psi^{(x')}_k\rangle\langle\Psi^{(x')}_k|)\\
\nonumber
& = & \sum_{j,k=0}^{d_A^2-1}\lambda_j^{(x)}\lambda_k^{(x')}|\langle\Psi^{(x)}_j|\Psi^{(x')}_k\rangle|^2.
\end{eqnarray}
For $x\neq x'$ we get that
\begin{eqnarray}
\nonumber
\sum_{j,k=0}^{d_A^2-1}\lambda_j^{(x)}\lambda_k^{(x')}|\langle\Psi^{(x)}_j|\Psi^{(x')}_k\rangle|^2 = 0,
\end{eqnarray}
but if we assume that $\rank(\rho_x) = d_A^2$, given that $\{|\Psi^{(x')}_k\rangle\}_{x=0,\dots,N-1}$ form a basis of $\mathcal{H}_{effective}$, the above sum cannot be null. At most, for a fixed value of $x$ and $x'$, $\rank(\rho_x) = d_A^2 - 1$ if $\rank(\rho_{x'}) = 1$. By extending this analysis to other values of $x'$, we conclude that $\rank(\rho_x) = 1$ for all $x$.

Because those states are pure, they can be written as
\begin{eqnarray}
|\Psi_x\rangle = \sum_{j=0}^{d_A-1}\eta_{j}^{(x)}|\psi_j^{(x)}\rangle\otimes|\phi_j^{(x)}\rangle.
\end{eqnarray}

Considering now the condition \eqref{eq: new condition} plus the unitary equivalence of the purifications \cite{Watrous2018}, we have that:
\begin{eqnarray}
\label{eq: pure preparations in the selftesting}
|\Psi_x\rangle = \sum_{j=0}^{d_A-1}\eta_{j}\left(U_x|\psi_j\rangle\right)\otimes|\phi_j\rangle,
\end{eqnarray}
in which $U_x^{\dagger}U_x=\Id$ and there is no loss of generality in setting $U_0=\Id$.

It holds that
\begin{eqnarray}
\nonumber
\sum_{x=0}^{d_A^2-1}|\Psi_x\rangle\langle\Psi_x| & = & \Id,
\end{eqnarray}
which implies that
\begin{multline}
\nonumber
\tr_A\left(\sum_{x=0}^{d_A^2-1}\sum_{j,k=0}^{d_A-1}\eta_j\eta_k(U_x|\psi_j\rangle\langle\psi_k|U_x^{\dagger})\otimes |\phi_j\rangle\langle\phi_k|\right) \\ =  d_A\Id_B.
\end{multline}
On the other hand we have that
\begin{multline}
\nonumber
\tr_A\left(\sum_{x=0}^{d_A^2-1}\sum_{j,k=0}^{d_A-1}\eta_j\eta_k(U_x|\psi_j\rangle\langle\psi_k|U_x^{\dagger})\otimes |\phi_j\rangle\langle\phi_k|\right) \\ = d_A^2\sum_{j=0}^{d_A-1}\eta_j^2|\phi_j\rangle\langle\phi_j|
\end{multline}
which can only happen if $\eta_j=\frac{1}{\sqrt{d_A}}$, which then concludes the proof.
\end{proof}

\section{N preparations and N(N-1)/2 dicotomic measurements}
\label{App: N preparations and N(N-1)/2}

\begin{result}
In a prepare and measure scenario with $N$ preparations $x$ and $N(N-1)/2$ dichotomic measurements $(y_1,y_2)$, for $y_1>y_2$, where $y_1,y_2\in\{0,\dots,N-1\}$, the set of probability distributions  is bounded by the inequality
\begin{eqnarray}
\label{eq:app general N(N-1)/2 measurements set}
V_N & \leq & \frac{N^2}{2}\left(1 - \frac{1}{\min (d_As, N)}\right),
\end{eqnarray}
where $d_A$ is the Hilbert-space dimension of the quantum system sent from Alice to Bob, $s$ is the Schmidt number of the quantum state shared between Alice and Bob, and
\begin{eqnarray}
V_N & = & \sum_{x>x'} \left|P(1|x,(x,x')) - P(1|x',(x,x')\right|^2.
\end{eqnarray}
Furthermore, if $s=d_A$ and $N< d_A^2$ or $N=cd_A^2$, for integer $c$, expression \eqref{eq:app general N(N-1)/2 measurements set} is tight.
\end{result}

\begin{proof}
First notice that if $s=1$ we have the case analyzed in reference \cite{Brunner2013}, and the result holds. We hereby analyze the remaining quantum cases, for $s>1$.

Let $\{\rho_x\}_{x=0,\dots,N-1}\subset S_s$ be the set of preparations for which relation \ref{eq: new condition} holds, and $\{M^{y,y'}_{b}\}_{b\in\{0,1\}}$ the measurements settings, so
\begin{eqnarray}
V_N & = & \sum_{x>x'}\left| \tr \left[(\rho_x - \rho_{x'})M^{(x,x')}_{b=1}\right]\right|^2.
\end{eqnarray}
Defining $\mbox{D}(\rho_x,\rho_{x'}) \coloneqq \frac{1}{2}||\rho_x - \rho_{x'}||_1$, it is known that \cite{nielsen2001}
\begin{eqnarray}
\nonumber
\mbox{D}(\rho_x,\rho_{x'}) & = & \underset{P\in \mathcal{P}(\mathcal{H}_A\otimes\mathcal{H}_B)}{\max}\tr\left(\left(\rho_x - \rho_{x'}\right)P\right),
\end{eqnarray}
where $\mathcal{P}(\mathcal{H}_A\otimes\mathcal{H}_B)$ is the set of positive operators that act on $\mathcal{H}_A\otimes\mathcal{H}_B$.
Clearly the following relation is always satisfied:
\begin{eqnarray}
\label{eq: app V_N in terms of trace distance}
V_N \leq \sum_{x > x'}\left|\mbox{D}(\rho_x,\rho_{x'})\right|^2.
\end{eqnarray}
Because of the triangle inequality, the right-hand side of the equation is a convex function of $\mbox{D}(\rho_x,\rho_{x'})$, which being a norm is also a convex function of $\rho_x - \rho_{x'}$, which can be seen as a map $F$, the domain of which is $\mbox{dom}(F)= \{\chi = \rho\otimes\sigma\; | \; \chi \in \mathcal{D}(\mathcal{H}_{A_1}\otimes\mathcal{H}_{B_1}\otimes\mathcal{H}_{A_2}\otimes\mathcal{H}_{B_2})\;\;\and\;\;\rho,\sigma\in\mathcal{D}(\mathcal{H}_A\otimes\mathcal{H}_B)\}$, given by:
\begin{eqnarray}
F(\chi) = \tr_{A_2B_2}(\chi) - \tr_{A_1B_1}(\chi).
\end{eqnarray}
It follows that $F$ is a linear map, and that $\mbox{dom}(F)$ is a convex set the extremal points of which are given by elements $\chi = \rho\otimes\sigma$ for which $\rho$ and $\sigma$ are pure states.

With this we can say that the right-hand side of equation \eqref{eq: app V_N in terms of trace distance} is a convex function defined in $\mbox{dom}(F)$ and thus has its maximal value at some extremal point in $\mbox{dom}(F)$. This implies that
\begin{eqnarray}
\label{eq: V_N in terms of the fidelity}
\nonumber
V_N & \leq & \sum_{x > x'}\left|\mbox{D}(|\Psi_x\rangle\langle\Psi_x|,|\Psi_{x'}\rangle\langle\Psi_{x'}|)\right|^2\\
\nonumber
    & = & \sum_{x > x'}\left|\left(1 - \left|\langle\Psi_x|\Psi_{x'}\rangle\right|^2\right)^{\frac{1}{2}}\right|^2\\
\nonumber
    & = & \sum_{x > x'}\left(1 - \left|\langle\Psi_x|\Psi_{x'}\rangle\right|^2\right)\\
\nonumber
    & = & \frac{N(N-1)}{2} - \sum_{x > x'}\left|\langle\Psi_x|\Psi_{x'}\rangle\right|^2\\
    & = & \frac{N(N-1)}{2} - \frac{1}{2}\left(\sum_{x , x'}\left|\langle\Psi_x|\Psi_{x'}\rangle\right|^2 - N\right).
\end{eqnarray}

Now, define $\Omega$ as follows:
\begin{eqnarray}
\Omega = \frac{1}{N}\sum_{x=0}^{N-1}|\Psi_x\rangle\langle\Psi_x|,
\end{eqnarray}
so the equation \eqref{eq: V_N in terms of the fidelity} can be expressed as
\begin{eqnarray}
\label{eq: V_N in terms of Omega}
V_N & \leq & \frac{N^2}{2} - \frac{N^2}{2}\tr (\Omega^2).
\end{eqnarray}
At this point, we recall that
\begin{eqnarray}
|\Psi_x\rangle = \sum_{j=1}^{s}\eta_j^{(x)}|\psi_j^{(x)}\rangle\otimes|\phi_j\rangle,
\end{eqnarray}
and condition \eqref{eq: new condition} plus the unitary equivalence of the purifications lead to
\begin{eqnarray}
|\Psi_x\rangle = \sum_{j=0}^{s}\eta_{j}\left(U_x|\psi_j\rangle\right)\otimes|\phi_j\rangle,
\end{eqnarray}
This implies that $\Omega \in \mathcal{D}(\mathcal{H}_{effective})$, i.e., $\Omega$ is a density operator acting on $\mathcal{H}_{effective}$,  where $\mathcal{H}_{effective}$ was defined in appendix \ref{App: Scenario of superdense coding} and has dimension $\dim (\mathcal{H}_{effective}) = d_As$. So we must have that
\begin{eqnarray}
\tr (\Omega^2) \geq \frac{1}{d_As},
\end{eqnarray}
which leads to
\begin{eqnarray}
\label{eq: app almost V_N}
V_N & \leq & \frac{N^2}{2}\left(1 - \frac{1}{d_As}\right).
\end{eqnarray}

Whenever $N\leq d_As$, $V_N$ we are working under the communication capacity of the channel, and $V_N$ always reaches its maximum algebraic value, so we can rewrite \eqref{eq: app almost V_N}:
\begin{eqnarray}
\label{eq: our witness}
V_N & \leq & \frac{N^2}{2}\left(1 - \frac{1}{\min (d_As, N)}\right).
\end{eqnarray}

Now we show that if $N< d_A^2$ or $N=cd_A^2$, for an integer $c$, the above expression is saturated using the measurements that optimally discriminate $|\Psi_x\rangle$ from $|\Psi_{x'}\rangle$ given that:
\begin{eqnarray}
\label{eq: app preparations N(N-1)/2 measurement settings} 
|\Psi_x\rangle = \frac{1}{\sqrt{d_A}}\sum_{j=0}^{d_A-1}U_x|j\rangle\otimes|j\rangle.
\end{eqnarray}
in which $\{U_x\}$ is a to be defined set of unitary operators acting on $\mathcal{H}_A$.
We prove that by showing that if the state defined as $\Omega$ is such that $\tr(\Omega^2) = \frac{1}{\min(d_A^2,N)}$, for the above preparations, then there exist measurements $\{M^{y,y'}_{b}\}_{b\in\{0,1\}}$, for $y_1>y_2$, with  $y_1,y_2\in\{0,\dots,N-1\}$ leading to a saturation of our witness, even though these are never specified.

From equation \eqref{eq: app preparations N(N-1)/2 measurement settings},
\begin{eqnarray}
\nonumber
\Omega & = & \frac{1}{Nd_A}\sum_{x=0}^{N-1}\sum_{j,k=0}^{d_A-1}\left(U_x|j\rangle\langle k|U_x^{\dagger}\right)\otimes|j\rangle\langle k|,
\end{eqnarray}
and:
\begin{multline}
\nonumber
\Omega^2  = \\ \frac{1}{N^2d_A^2}\sum_{x,x'=0}^{N-1}\sum_{j,k,m=0}^{d_A-1}\left(U_x|j\rangle\langle k|U_x^{\dagger}U_{x'}| k\rangle\langle m|U_{x'}^{\dagger}\right)\otimes|j\rangle\langle m|.
\end{multline}
Straight forward calculations lead to:
\begin{eqnarray}
\nonumber
\tr (\Omega^2) & = & \frac{1}{N^2d_A^2}\sum_{x,x'=0}^{N-1}\tr\left(U_{x'}^{\dagger}U_{x}\right)\tr\left(U_{x}^{\dagger}U_{x'}\right)
\end{eqnarray}
Fixing the set $\{U_x\}$  as the set of Weyl operators $\{W_x\}_{x=0,\dots ,d_A^2-1}$ acting on $\mathcal{H}_A$, and letting $c=\left\lfloor\frac{N}{d_A^2}\right\rfloor$, i.e., $c$ is the integer part of $\frac{N}{d_A^2}$, we have
\begin{eqnarray}
N = cd_A^2 + N \bmod d_A^2.
\end{eqnarray}
Define $\mathcal{N} = N \bmod d_{A^{2}}$. Now, we are going to divide the set of preparations $\{0,\dots,N-1\}$ into $d_A^2$ groups. If $x$ and $x'$ are in the same group, then $|\Psi_x\rangle = |\Psi_{x'}\rangle$. There will be $ \mathcal{N}$ groups with $c + 1$ members and $N -\mathcal{N}$ groups with $c$ members. Each group is defined by a Weyl operator $W_X$ so we have that
\begin{eqnarray}
\nonumber
\tr (\Omega^2) & = & \frac{1}{N^2d_A^2}\sum_{x,x'=0}^{N-1}d_A^2\delta_{X,X'}\\
\nonumber
               & = & \frac{1}{N^2}\left(\sum_{x=0}^{\mathcal{N} -1}(c + 1) + \sum_{x=N\bmod d_A^2}^{N-1}c\right)\\
\nonumber
               & = & \frac{1}{N^2}\left((c + 1)\mathcal{N}\right.\\
               \nonumber
               &   & \left.+ c(N - \mathcal{N})\right)\\
\nonumber
               & = & \frac{1}{N^2}\left(\mathcal{N} + cN\right).
\end{eqnarray}
Clearly the above expression is $\frac{1}{N}$ if $N<d_A^2$ (in this case $c=0$ and $\mathcal{N} = N$), and $\frac{1}{d_A^2}$ if $N=cd_A^2$, for integer $c$ (here we have that $\mathcal{N} = 0$). This exactly saturates the bound of equation \eqref{eq: our witness}.
\end{proof}
\end{document}